\documentclass{sig-alternate-05-2015}

\usepackage{epstopdf}
\usepackage{algorithm} 
\usepackage{algorithmic}
\usepackage{tabularx}
\usepackage{float}

\usepackage{graphicx}
\usepackage{array}
\usepackage{listings}
\usepackage{booktabs}
\usepackage{textcomp}

\usepackage{amsmath}
\usepackage{color}

\begin{document}

\newtheorem{theorem}{Thm.}
\newtheorem{proposition}{Prop.}
\newtheorem{corollary}{Cor.}
\newtheorem{definition}{Def.}



%

\title{A Random Sample Partition Data Model for Big Data Analysis}
%
%
%
%
%

\numberofauthors{4} 
%
\author{
%
%
 Salman Salloum \space \space \space Yulin He \space \space \space Joshua Zhexue Huang\titlenote{Corresponding Author} \space \space \space  Xiaoliang Zhang \space \space \space Tamer Z. Emara\\
 Chenghao Wei \space \space \space Heping He
 \\
       \affaddr{Big Data Institute, College of Computer Science and Software Engineering, Shenzhen University,}\\
       \affaddr{Shenzhen 518060, Guangdong, China}\\
       \email{\{ssalloum, yulinhe, zx.huang, zhangxlassz, tamer, chenghao.wei\}@szu.edu.cn} \\
       \email{heping.he@outlook.com}
}

\maketitle
\begin{abstract}

Big data sets must be carefully partitioned into statistically similar data subsets that can be used as representative samples for big data analysis tasks. In this paper, we propose the random sample partition (RSP) data model to represent a big data set as a set of non-overlapping data subsets, called RSP data blocks, where each RSP data block has a probability distribution similar to the whole big data set. Under this data model, efficient block level sampling is used to randomly select RSP data blocks, replacing expensive record level sampling to select sample data from a big distributed data set on a computing cluster. We show how RSP data blocks can be employed to estimate statistics of a big data set and build models which are equivalent to those built from the whole big data set. In this approach, analysis of a big data set becomes analysis of few RSP data blocks which have been generated in advance on the computing cluster. Therefore, the new method for data analysis based on RSP data blocks is scalable to big data.

%
%
\end{abstract}

\section{Key insights}

\begin{itemize}
\item Big data analysis requires not only computationally efficient, but also statistically effective approaches.
\item A random sample partition (RSP) data model for big data can be used efficiently and effectively in building ensemble models, estimating statistics, and other big data analysis tasks on a computing cluster.
\item Experiments show that a sample of RSP data blocks from an RSP data model can be used to get approximate analysis results equivalent to those using the whole data set.
\item In addition to the scalability advantage, adopting this approach saves time and computing resources.
\end{itemize}

\section{Introduction}

Big data analysis is a challenging problem in many application areas especially with the ever increasing volume of data. In this regard, divide-and-conquer is used as a common strategy in current big data frameworks to distribute both data and computation on computing clusters. A big data set is divided into smaller data blocks and distributed on the nodes of a computing cluster so that data-parallel computations are run on these blocks. However, analyzing a big data set all at once may require more computing resources than available ones on the computing cluster. Moreover, current frameworks require new data-parallel implementations of traditional data mining and analysis algorithms. In order to enable efficient and effective big data analysis when data volume goes beyond the available computing resources, a different approach is required which considers computational, as well as statistical, aspects of big data on both data management and analysis levels. Such an approach is essential for investigating a key research problem nowadays: should the full set of data be used to find properties and reveal valuable insights from big data or a subset of this data be good enough?

Although the existing mainstream big data frameworks (e.g., Hadoop's MapReduce \cite{ISI:000251994700031}, Apache Spark \cite{Zaharia2012}) run data-parallel computations over distributed data blocks on computing clusters, current data partitioning techniques do not consider the probability distributions of data in these blocks. Sequentially chunking a big data set into small data blocks does not guarantee that each block is a random sample in case that data is not randomly ordered in the big data set. In such case, using data blocks directly to estimate statistics and build models may lead to statistically incorrect or biased results. Furthermore, classical random sampling techniques, which require a full scan of the data set each time a random sample is generated, are no longer effective with the increasing volume of big data sets stored in distributed systems \cite{export:159275}. Thus, partitioning a big data set into small data subsets (i.e., data blocks), each being a random sample of the whole data set, is a fundamental operation for big data analysis. These data blocks can be used to estimate statistics and build models, especially when analyzing big data sets requires more than the available resources in order to meet specific application requirements \cite{goiri2015approxhadoop}\cite{Gantz2011}. Therefore, it is necessary to develop statistically-aware data partitioning methods which enable effective and efficient usage of data blocks to fulfill the statistical and scalability requirements of big data analysis tasks.

Multivariate data is a common form of data in many application areas. Let $\mathbb{D}$$= \left\{ {{{\rm{x}}_1},{{\rm{x}}_2}, \cdots ,{{\rm{x}}_N}} \right\}$ be a multivariate data set of $N$ records where $N$ is too big for statistical analysis of $\mathbb{D}$ on a single machine. Each record is depicted with $M$ attributes or features, i.e., ${{\rm{x}}_i} = \left( {{x_{i1}},{x_{i2}}, \cdots ,{x_{iM}}} \right)$ for any $i \in \left\{ {1,2, \cdots ,N} \right\}$. In order to make a statistically-aware data partitioning, we propose the \textbf{random sample partition} (RSP) data model to represent $\mathbb{D}$ as a set of non-overlapping data blocks ( called RSP data blocks) where each block itself is a random sample. If the records in $\mathbb{D}$ are independently and identically distributed (i.i.d), we call $\mathbb{D}$ a randomized data set (i.e., the order of records in $\mathbb{D}$ is random). Our previous empirical study \cite{Salloum:2016:EAA:3006299.3006306} showed that data sets are naturally randomized in many application areas and thus can be directly represented as RSP data models using the current data partitioning techniques. In case that the data is not randomized, scalable algorithms are developed to randomize big data sets on computing clusters.

The RSP data blocks in an RSP can be directly drawn as random samples of the whole big data set in data exploratory and analysis tasks. Since the RSP data blocks are generated in advance and stored on the computing cluster, randomly selecting a set of RSP data blocks is much efficient than sampling a set of records from a distributed big data file because the full scan on the whole file is no longer needed.
In this article, we show that a small sample of few RSP data blocks from an RSP is enough to build models and compute the statistical estimates which are equivalent to those calculated from the whole big data set. We propose the asymptotic ensemble learning framework for big data analysis which depends on ensemble methods as a general approach for building block-based ensemble models from RSP data blocks. With this framework, results can be improved incrementally without the need to load and analyze the whole big data set all at once. This approach can be generalized for statistics estimation by defining appropriate ensemble functions (e.g., averaging the means from different RSP data blocks). 

In this article, we introduce the RSP data model of a big data set and show how RSP data blocks are essential for efficient and effective big data analysis. We also discuss how this new RSP data model can be employed for different data analysis tasks using the asymptotic ensemble learning framework. Finally, we summarize the implications of this new model and conclude this article with some of our current works.

\section{Distributed Data-Parallel Computing}
As data volume in different application areas goes beyond the petabyte scale, divide-and-conquer is taken as a general strategy to process big data on computing clusters considering the recent advancements in distributed and parallel computing technology. In this strategy, a big data file is chunked into small non-overlapping data blocks and distributed on the nodes of a computing cluster using a distributed file system such as Hadoop distributed file system (HDFS) \cite{hdfs2010}. Then, data-parallel computations are run on these blocks considering data locality. After that, intermediate results from processed blocks are integrated to produce the final result of the whole data set. This is usually done using the MapReduce computing model \cite{ISI:000251994700031} adopted by the mainstream big data frameworks such as Apache Hadoop \footnote{http://hadoop.apache.org/}, Apache Spark \footnote{https://spark.apache.org/}, and Microsoft R Server \footnote{https://www.microsoft.com/en-us/cloud-platform/r-server}.

As a unified engine for big data processing, Apache Spark has been adopted in a variety of applications in both academia and industry \cite{Zaharia:2016:ASU:3013530.2934664}\cite{Salloum2016} as a new generation engine after Hadoop's MapReduce. It uses a new data abstraction and in-memory computation model, the resilient distributed datasets (RDDs) \cite{Zaharia2012}, where collections of objects (e.g., records) are partitioned across a cluster, kept in memory and processed in parallel. Similarly, Microsoft R Server addresses the in-memory limitations of the open source statistical system, R, by adding parallel and chunk-wise distributed processes across multiple cores and nodes. It comes with the proprietary eXternal data frame (XDF) and a framework for parallel external memory algorithms (PEMAs) of statistical analysis and machine learning. Both Apache Spark and Microsoft R Server operate on data stored on HDFS. A fundamental operation is importing such data into XDF format or RDDs, and then running data-parallel operations.

Although the current frameworks employ the data-parallel model to run scalable algorithms on computing clusters, analyzing a whole big data set may exceed the available computing resources. There are some technical solutions for this problem such as loading and processing blocks in batches according to the available resources, but this still requires analyzing each block in the data set which leads to longer computation time. Furthermore, current data partitioning techniques simply cut data sets into blocks without considering data probability distributions in these blocks. This can lead to statistically incorrect or biased results in some data analysis tasks. We argue that solving big data analysis problems requires solutions which not only consider the computational aspects of big data, but also the statistical ones. To address this issue, we propose the RSP data model.

\section{RSP of Big Data}

 The RSP data model is a new model to represent a big data set as a set of RSP data blocks. It depends on two fundamental concepts: random sample and partition of a data set. First, we define these basic concepts and show how they are applied to the RSP data models of big data sets. Then, we present a formal definition of the RSP data model. For convenience of discussion, we do not consider the set anisotropy.

\paragraph{Random Sample of a Big Data Set}

Sampling is an essential technique for big data analysis when the volume of data goes beyond the available computing resources. Random samples are widely used in statistics to explore the statistical properties and distributions of big data, calculate the statistics estimates and build regression and classification models. In big data analysis, we define a random sample of a big data set as follows:\\

		\noindent
		\textbf{Definition 1 (Random Sample)}: Let $\mathbb{D}$ $=$ $\left\{ {{{\rm{x}}_1}} \right.,$ ${{\rm{x}}_2},$ $\cdots,$ $\left. {{{\rm{x}}_N}} \right\}$ be a big data set of $N$ records. Let $D_n$ be a subset of $\mathbb{D}$ containing $n$ records chosen from $\mathbb{D}$ using a random process. $D_n$ is a random sample of $\mathbb{D}$ if
 $$E[\tilde{F}_n(x)]=F(x),\,\,{\rm for \,\,}n \leq N$$
 where $\tilde{F}_n(x)$ and $F(x)$ denote the sample distribution functions of $\rm{D}_n$ and $\mathbb{D}$, respectively. $E[\tilde{F}_n(x)]$ denotes the expectation of $\tilde{F}_n(x)$. 

According to the law of large numbers, we consider that a big data set is a random sample of the population in a certain application domain. For example, a big data set of customers is a random sample of the customer's population in a company. Thus, we can use such data set to estimate the distribution of all customers in the company. For big data analysis,
a random sample of a big data set $\mathbb{D}$ is often used to investigate the statistical properties of $\mathbb{D}$ when $\mathbb{D}$ is too big to be analyzed using the available computing resources. A random sample is taken from $\mathbb{D}$ using a random sampling process as stated in Lemma 1. However, if $\mathbb{D}$ is distributed on a computing cluster, taking a random sample from $\mathbb{D}$ itself is an computationally expensive process because a full scan of $\mathbb{D}$ needs to be conducted on the distributed nodes of the computing cluster. To avoid the random sampling process on $\mathbb{D}$ in big data analysis, we can generate a set of random samples from $\mathbb{D}$ in advance and save these random samples to be used in analysis of $\mathbb{D}$. This idea is materialized in the RSP data model.


\paragraph{Partition of a Big Data Set}
On distributed file systems, a big data set $\mathbb{D}$ is divided into small data blocks which are distributed on the nodes of a computing cluster. Each block contains a subset of records from $\mathbb{D}$. From a mathematical point of view, this set of blocks forms a partition of $\mathbb{D}$. \textit{In mathematics, a partition of a set is a grouping of the set's elements into non-empty subsets, in such a way that every element is included in one and only one of the subsets\footnote{https://en.wikipedia.org/wiki/Partition\_of\_a\_set}}. \\

\noindent
\textbf{Definition 2 (Partition of Data Set)}: Let $\mathbb{D}$ $=$ $\left\{ {{{\rm{x}}_1}} \right.,$ ${{\rm{x}}_2},$ $\cdots,$ $\left. {{{\rm{x}}_N}} \right\}$ be a data set containing $N$ objects. Let $\mathbb{T}$ be an operation which divides $\mathbb{D}$ into a set of subsets ${\rm{T}} = \left\{ {{{\rm{D}}_1},} \right.{{\rm{D}}_2}, \cdots ,\left. {{{\rm{D}}_K}} \right\}$. ${\rm{T}}$ is called a \textit{ partition} of data set $\mathbb{D}$ if \\
(1) $\bigcup\limits_{k = 1}^K {{{\rm{D}}_k}} = $ $\mathbb{D}$;\\
(2) ${{\rm{D}}_i} \cap {{\rm{D}}_j} = \emptyset$, when $i,j \in \left\{ {1,2, \cdots ,K} \right\}$ and $i \ne j$.\\
Accordingly, $\mathbb{T}$ is called a partition operation on $\mathbb{D}$ and each ${{{\rm{D}}_k}}(k=1,2,\cdots,K)$ is called a data block of $\mathbb{D}$. \\


According to this definition, many partitions can be generated from a data set $\mathbb{D}$. For example, an HDFS file is a particular partition of $\mathbb{D}$ generated by partitioning operation $\mathbb{T}$ to sequentially cutting the original data file $\mathbb{D}$ into data blocks $ \left\{ {{{\rm{D}}_1},} \right.{{\rm{D}}_2}, \cdots ,\left. {{{\rm{D}}_K}} \right\}$. Even with the same partitioning operation, different partitions can be generated using different partitioning parameters such as the size of each block or the number of records in each block. However, a key issue with this sequential partitioning is that data blocks may not have similar statistical properties as the big data set. In fact, it is theoretically possible and practically required to find alternatives for generating a partition of a data set which satisfies certain application requirements or holds some statistical properties, e.g., each data block is a random sample of $\mathbb{D}$.

\paragraph{Random Sample Partition of a Big Data Set}

We define the Random Sample Partition (RSP) to represent a big data set as a family of non-overlapping random samples.\\

\noindent
\textbf{Definition 3 (Random Sample Partition)}: Let $\mathbb{D}$ $=$ $\left\{ {{{\rm{x}}_1}} \right.,$ ${{\rm{x}}_2},$ $\cdots,$ $\left. {{{\rm{x}}_N}} \right\}$ be a big data set which is a random sample of a population and assume $F(x)$ is the sample distribution function (\textit{s.d.f.}) of $\mathbb{D}$. Let $\mathbb{T}$ be a partition operation on $\mathbb{D}$ and ${\rm{T}}$ $=$ $\left\{ {{{\rm{D}}_1}} \right.{\rm{,}}$ ${{\rm{D}}_2},$ $\cdots,$ $\left. {{{\rm{D}}_K}} \right\}$ be a partition of $\mathbb{D}$ accordingly. ${\rm{T}}$ is called a \textit{random sample partition} of $\mathbb{D}$ if
$$E[\tilde{F}_k(x)]=F(x)\,\,\,{\rm for\,\,\, each}\,\,\, k=1,2,\cdots,K,$$
where $\tilde{F}_k(x)$ denotes the sample distribution function of $\rm{D}_k$ and $E[\tilde{F}_k(x)]$ denotes its expectation. Accordingly, each $\rm{D}_k$ is called an RSP data block of $\mathbb{D}$ and $\mathbb{T}$ is called an RSP operation on $\mathbb{D}$. In the next section, we discuss the partitioning process to generate RSPs from big data sets.



\section{Two-Stage Data Partitioning}
\label{partitioning_algorithm}

Given a big data set in HDFS, a partitioning algorithm is required to convert the HDFS file into an RSP data model of $K$ RSP data blocks, each with $n$ records, which are also stored in HDFS. In this case, the partitioning algorithm consists of two main steps: data chunking and data randomization as discussed below. The pseudo code is given in Algorithm \ref{alg:random_partitioner}.

\begin{itemize}
\item Data Chunking: $\mathbb{D}$ is divided into $P$ data blocks. We call these blocks the original data blocks which are not necessarily randomized. In the current big data frameworks, this operation is straightforward and available.
	
\item Data Randomization: a slice of $\delta$ records is randomly selected without replacement from each of the original $P$ data blocks to form a new RSP data block. First, each original data block is randomized locally. Then, the randomized block is further chunked into $K$ sub-blocks, each with $\delta$ records. After that, a new RSP block is created by selecting one sub-block from each of the randomized original blocks and combining these selected sub-blocks together. The last step is repeated $K$ times to produce the required number of RSP data blocks where each RSP data block has $n=K\delta$ records. 
\end{itemize}

The number of records in a slice $\delta$ can be determined depending on $K$ and $n$ (e.g., $\delta$ = $n/K$ to distribute data evenly over all blocks). The number of RSP data blocks $K$, or the number of records in an RSP data block $n$, is selected depending on both the available computing resources and the target analysis tasks so that a single RSP data block can be processed efficiently on one node (or core). 

%

\begin{algorithm}[t]
	\caption{Partitioning Algorithm}
	\label{alg:random_partitioner}
	\begin{algorithmic}
		\STATE {\bfseries Input:}
		\STATE - $\mathbb{D}$: big data set;
		\STATE - $P$: number of data blocks of $\mathbb{D}$;
		\STATE - $n$: number of records in an RSP data block;
		\STATE - $K$: number of required RSP data blocks;
		\STATE {\bfseries Method:}
		\STATE {$\delta = n/K$};
		\STATE Divide $\mathbb{D}$ into $P$ data blocks;
		\FOR  {$i=1$ to $P$}
		\STATE Randomize block ${\rm{D}}_i$;
		\STATE Sequentially cut each randomized ${\rm{D}}_i$ into $K$ sub-blocks, each with $\delta$ records;
		\ENDFOR
		\FOR  {$k=1$ to $K$}
		\STATE $\rm{D}_k = \phi$;
		\FOR  {$i=1$ to $P$}
		\STATE Select one sub-block from $\rm{D}_i$ without replacement;
		\STATE Append the sub-block to $\rm{D}_k$;
		\ENDFOR	
		\STATE Save $D_k$ as an RSP data block;
		\ENDFOR
		\STATE {$\rm{T} =  \left\{ {{{\rm{D}}_1},} {{\rm{D}}_2}, \cdots , {{{\rm{D}}_K}} \right\}$};
		\STATE {\bfseries Output:}
		\STATE - $\rm{T}$: a set of RSP data blocks;			
	\end{algorithmic}
\end{algorithm}

\begin{figure}[!htb]
	\begin{center}
		\includegraphics[width=8cm]{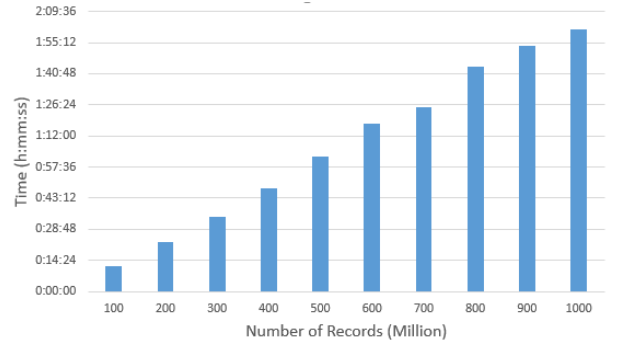}
		\caption{Time for creating RSPs from synthesized numerical datasets with 100 features using Apache Spark on a computing cluster of 5 nodes (each node has 24 cores, 128 GB RAM and 12.5 TB disk storage). The algorithm was run on 10 different sizes of data, from 100 million records (1000 blocks) to 1 billion records (10000 blocks). The storage size is approximately 100 GB for 100 million records and 1 TB for 1 billion records. }
		\label{fig:synthesized_randomization_time}
	\end{center}
\end{figure}

A prototype of this algorithm was implemented using the Apache Spark's RDD API. As a preliminary test on a small cluster of 5 nodes (each node has 24 cores, 128 GB RAM and 12.5 TB disk storage), Figure (\ref{fig:synthesized_randomization_time}) shows the partitioning time for synthesized numerical data. We can see how the partitioning time increases almost linearly with the number of records in the data set. This shows the scalability of such partitioning algorithms. In this experiment, the number of RSP data blocks is the same as the number of original blocks where each block contains 100,000 records. However, we also found that the partitioning time did not vary much when the number of RSP data blocks is different from the number of original blocks. Thus, this kind of algorithms can be used to fulfill different requirements as the number of records in a block is an essential factor for some analysis tasks.

\section{Probability Distributions of RSP data blocks} 
The key idea behind an RSP is partitioning a big data set into RSP data blocks which can be used directly as random samples of the whole data set. Thus, it is essential to investigate the statistical properties of these RSP data blocks. We used hypothesis testing and exploratory data analysis in our previous empirical studies \cite{Salloum:2016:EAA:3006299.3006306} to compare RSP data blocks. We discuss here the underlying theory of using the RSP data blocks as random samples of the whole data set.

\begin{figure*}[!htb]
	\begin{center}
		\includegraphics[width=\textwidth]{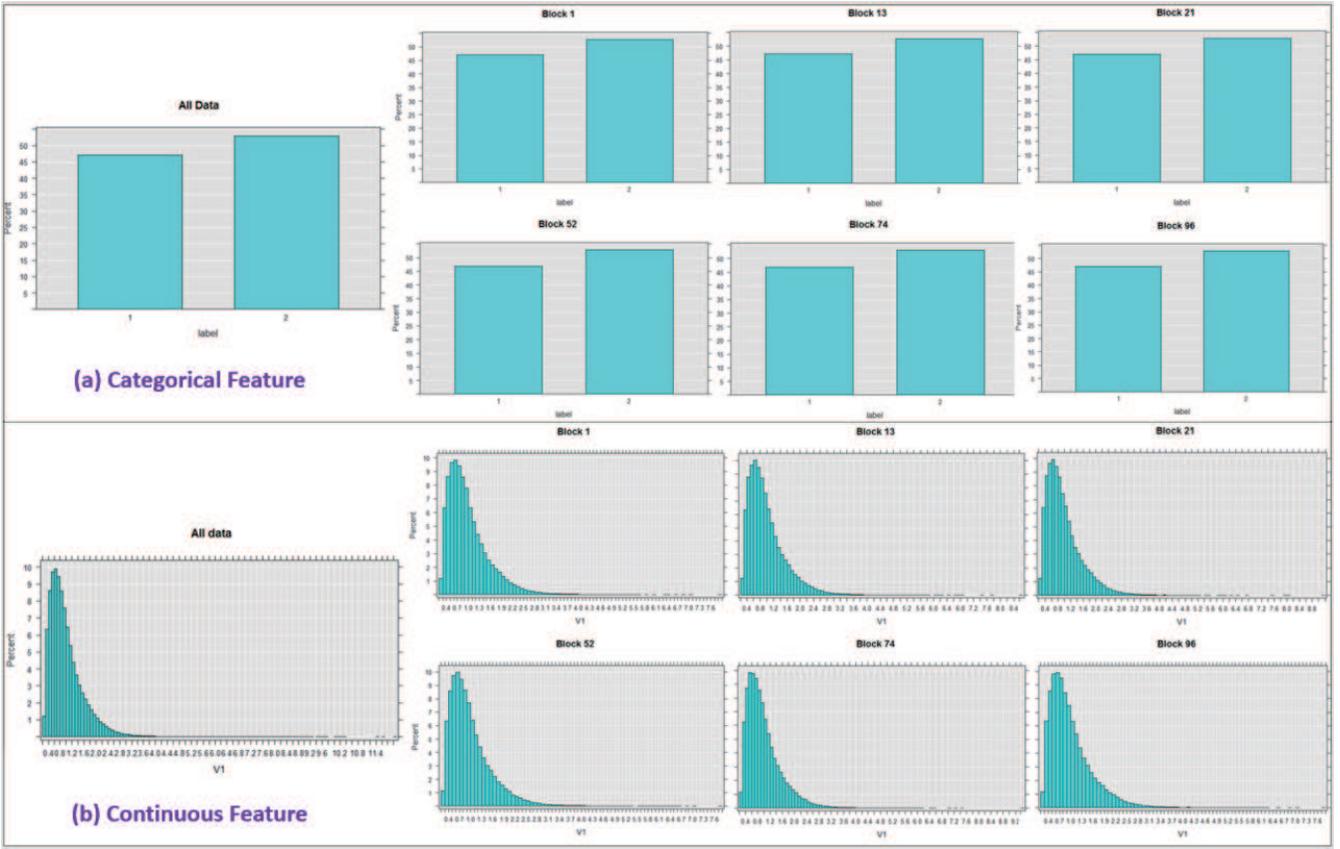}
		\caption{Probability distribution in data blocks and the whole data set.}
		\label{fig:higgs_data_distribution}
	\end{center}
\end{figure*}

Each block in an RSP should follow a data distribution similar to the whole data set.
For a categorical feature, records that belong to the same category should be evenly distributed over all blocks. A good example is the label feature in a classification task. For example, Figure (\ref{fig:higgs_data_distribution}.a) shows the distribution of a label feature (from HIGGS data set\footnote{https://archive.ics.uci.edu/ml/datasets/HIGGS}) in the whole data set and some randomly selected RSP data blocks. This also applies to continuous features as shown in Figure (\ref{fig:higgs_data_distribution}.b). In addition, the similarity between RSP data blocks and samples by simple random sampling, or the whole data set if possible, can be computed using quantity measures such as MMD measure \cite{Gretton:2012:KTT:2188385.2188410}. 

Selecting a block from an RSP of  $\mathbb{D}$ is equivalent to drawing a random sample directly from the big data set $\mathbb{D}$. If $\mathbb{D}$ is a randomized data set, cutting a block of $n$ records is equivalent to randomly selecting these $n$ records. This also applies when partitioning a non-randomized data set using Algorithm \ref{alg:random_partitioner} because each RSP data block is formed from randomly selected records from all the original blocks. Thus, each RSP data block from an RSP data model is equivalent to a simple random sample from $\mathbb{D}$. The underlying theory of RSP data blocks is shown in \textbf{Lemma 1} and \textbf{Theorem 1}.\\

\noindent
\textbf{Lemma 1}:
{Suppose $D=\{x_1,x_2,\cdots,x_{K\delta}\}$ is a data set. Randomly choose a permutation of the sequence $1,2,\cdots,K\delta$ which is denoted as $\tau=\{\tau_1,\tau_2,\cdots,\tau_{K\delta}\}$. For each $i=1,\cdots,K$, set $D_i=\{x_{\tau_{\delta(i-1)}+1},x_{\tau_{\delta(i-1)}+2},\cdots,x_{\tau_{\delta i}}\}$, then each $D_i$ is an RSP data block of $D$. Below, we prove this for one dimensional data. It also applies to high dimensional cases with minor modifications. \\

\noindent
\textbf{\emph{Proof}}: Suppose the sample distribution function of $D$ and $D_i$ are $F(x)$ and $F_i(x)$ $(i=1,2,\cdots,K)$ respectively. For any real number $x$, let $M$ denote the number of the records in $D$ whose values are not greater than $x$. Then, it is easy to show $M=K\delta\cdot F(x)$. If $s_i$ denotes the number of records in $D_i$ whose values are not greater than $x$, then
\begin{align*}
  E(s_i)&=\sum_{j=1}^{\delta}j\cdot P\{s_i=j\}=\sum_{j=1}^{\delta}j\frac{C^j_M\cdot C^{\delta-j}_{K\delta-M}}{C^{\delta}_{K\delta}}\\
        &=\frac{M}{C^{\delta}_{K\delta}}\cdot C^{^{\delta-1}}_{_{K\delta-1}}=\delta\cdot F(x),
\end{align*}
So, the expectation of $F_i(x)$ is
$$E[F_i(x)]=\frac{E(s_i)}{\delta}=F(x).$$
Thus, each $D_i$ is an RSP data block of $D$.\\
}

\textbf{Theorem 1}:
{
Suppose data set $A$ has $N_1$ records and data set $B$ has $N_2$ records. If $A_1$ is an RSP data block of $A$ with records $n_1$ and $B_1$ is an RSP data block of $B$ with records $n_2$ , then $A_1 \bigcup B_1$ is an RSP data block of $A \bigcup B$ as $\frac{n_1}{n_2}=\frac{N_1}{N_2}$.}\\

\textbf{\emph{Proof}}: Suppose the sample distribution functions of data blocks $A$, $B$, $A_1$ and $B_1$ are $F_1(x)$, $F_2(x)$, $\tilde{F}_1(x)$ and $\tilde{F}_2(x)$, respectively. We have $E[\tilde{F}_1(x)]=F_1(x)$ and $E[\tilde{F}_2(x)]=F_2(x)$.
For any real number $x$, the number of records of $A_1 \bigcup B_1$ whose values are not greater than $x$ is $n_1\tilde{F}_1(x)+n_2\tilde{F}_2(x)$. Therefore, the sample distribution function of $A_1 \bigcup B_1$ is:
$$
\tilde{F}(x)=\frac{n_1\tilde{F}_1(x)+n_2\tilde{F}_2(x)}{n_1+n_2}.
$$
Similarly, the sample distribution function of $A \bigcup B$ is:
$$
F(x)=\frac{N_1F_1(x)+N_2F_2(x)}{N_1+N_2}.
$$
The expectation of $\tilde{F}(x)$ is
\begin{align*}
 E[\tilde{F}(x)]&=E[\frac{n_1\tilde{F}_1(x)+n_2\tilde{F}_2(x)}{n_1+n_2}]\\
                &=\frac{n_1E[\tilde{F}_1(x)]+n_2E[\tilde{F}_2(x)]}{n_1+n_2}\\
                &=\frac{\frac{n_1}{n_2}F_1(x)+F_2(x)}{\frac{n_1}{n_2}+1}=\frac{\frac{N_1}{N_2}F_1(x)+F_2(x)}{\frac{N_1}{N_2}+1}\\
                &=F(x).
\end{align*}
Therefore, $A_1 \bigcup B_1$ is an RSP data block of $A \bigcup B$.

\section{Block Level Sample: A Sample of Random Sample Data Blocks}
As each block in an RSP data model is a random sample of big data set $\mathbb{D}$, the RSP data model can be used as a reduced representative space to directly draw samples of RSP data blocks instead of drawing samples of individual records from $\mathbb{D}$. In contrast to the classical record level sampling, we call this sampling method as block level sampling where blocks are sampled, without replacement and with equal probability. Thus, a sample of RSP data blocks is a set of non-overlapping random samples from $\mathbb{D}$. This is more efficient especially when $K << N$ because it avoids the full scan of $\mathbb{D}$ each time a random sample is needed.\\

\noindent
\textbf{Definition 4 (Block Level Sample)}: Let ${\rm{T}}$ be an RSP of a big data set $\mathbb{D}$. ${\rm{S}} = \left\{ {{{\rm{D}}_1},} \right.{{\rm{D}}_2}, \cdots ,\left. {{{\rm{D}}_g}} \right\}$, where $g < K$, is a block level sample with RSP data blocks randomly selected from ${\rm{T}}$ without replacement and with equal probability.

The block-sampling operation is called separately in each analysis process so that samples of RSP data blocks are selected without replacement, i.e., without repeating a block neither in the same sample nor in other samples in the same analysis process. The number of RSP data blocks $g$ in a block sample depends on the available computing resources so that each selected block can be loaded and analyzed locally on one node or core. In such case, a sample of RSP data blocks is analyzed in one batch in a perfectly parallel manner. This sampling process can be refined to select blocks depending on the availability of nodes on a computing cluster which can lead to better scheduling algorithms. 

Block level sampling can be employed for designing data analysis pipelines, for example, analyzing big data in batches (i.e., each batch uses one sample of RSP data blocks) to stepwise collect and improve statistical models and estimates rather than trying to load the data set all at once. Further statistical investigation can be done on the selected RSP data blocks. For example, exploratory data analysis techniques can be used to visualize and compare data distributions between RSP data blocks and classical random samples. Quantitative measures can be also used, such as the maximum mean discrepancy (MMD) measure \cite{Gretton:2012:KTT:2188385.2188410} for the similarity between data distributions and Hotelling's T-Square test for the difference between means of features. This kind of statistical testing is fundamental for many data analysis tasks. In the following sections, we describe how block level samples can be used effectively and efficiently for estimating statistics and building ensemble models.

\section{Estimation from a Sample of RSP Data Blocks}
The method of bootstrapping multiple samples from a given data set to evaluate statistical estimates is widely used in statistics. However, traditional bootstrap method is not suitable for a distributed big data set because of the high computational and storage costs. One approach for reducing these costs is drawing samples of small sizes such as the bag of little bootstraps \cite{Kleiner:2012tm} to calculate the average estimates from those small samples. However, this still requires scanning the whole data set each time a sample is needed. In addition, extra storage space to store these samples is also inevitable in most cases. Instead, storing a big data set in an RSP data model can avoid both the repeated random sampling and extra storage costs. This gives a chance, not only to directly estimate statistics and sample distributions from RSP data blocks, but also to turn the focus into improving the quality of estimations because the whole data set is stored, by default, as a collection of random samples.

Estimated statistics from individual RSP data blocks of an RSP data model can be combined (e.g., by averaging) in a single estimation which is generally better than any of the individual ones. For example, Figure (3) shows the average means of 4 features in HIGGS data. Each value is an average of the estimated means from blocks used up to that point. We can see the error of the means is not significant even in the first several batches and the estimated mean values converge to the true mean value while adding more blocks. In a similar way, Figure (4) shows the estimated standard deviations for the same features in HIGGS data. As we can see, samples of RSP data blocks from an RSP data model can be used efficiently to estimate the statistics of a big data set. This approach can be generalized to other statistical analysis tasks such as  building classification or regression models. In the following section, we show how a sample of RSP data blocks can be used effectively for building ensemble models.

		\begin{figure}[!htb]
			\centering
				\label{fig:higgs_mean}
				\includegraphics[width=9cm]{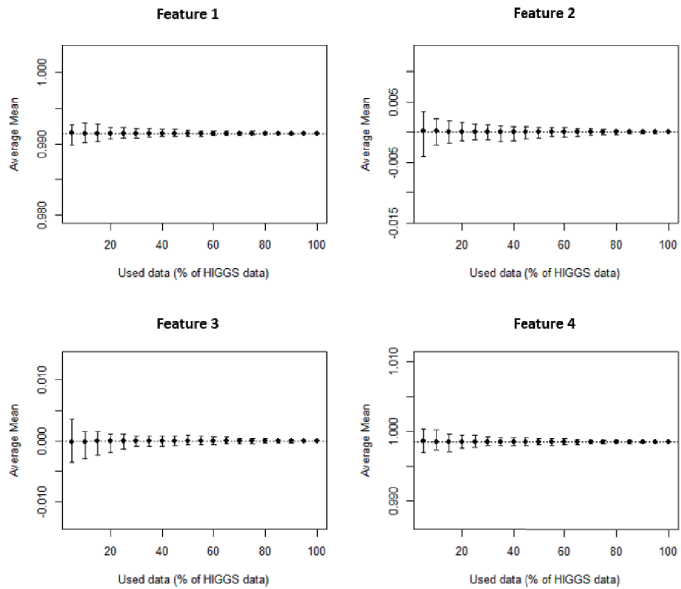}
			\caption{Block level estimation of means of 4 features in HIGGS data using 5 RSP data blocks. Each point represents the estimated value after each batch (averaged from 100 runs). The dotted line represents the true value calculated from the entire HIGGS data.}
		\end{figure}
		\begin{figure}[!htb]
				\label{fig:higgs_stddev}
				\includegraphics[width=9cm]{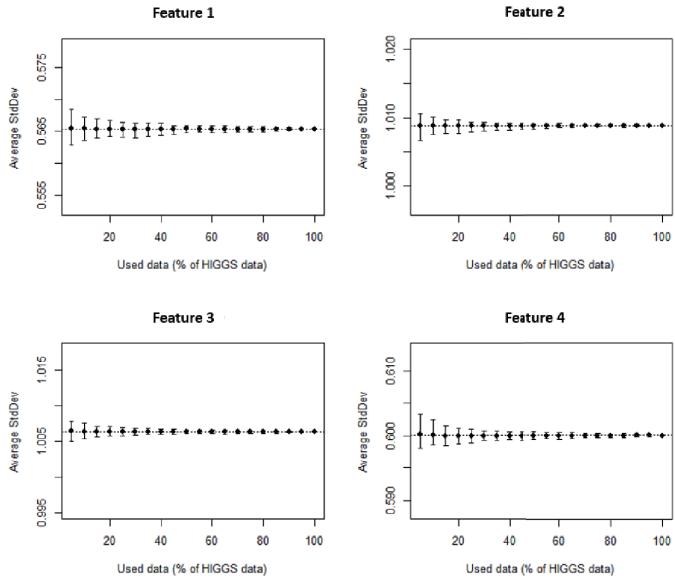}
			\caption{Block level estimation of standard deviations of 4  features in HIGGS data using 5 RSP data blocks. Each point represents the estimated value after each batch (averaged from 100 runs). The dotted line represents the true value calculated from the entire HIGGS data.}
		\end{figure}

\section{Asymptotic Ensemble Learning Framework}
The RSP data blocks of an RSP data model can be used as component data sets to build ensemble models for different tasks (e.g., classification, regression, clustering). Ensemble methods use multiple base models built with the same or different algorithms from different component data sets. The results from base models are combined in a single ensemble model which generally outperforms each of the individual ones \cite{Aggarwal:2015:DMT:2778285}.
\begin{figure*}[!htb]
	\begin{center}
		\includegraphics[width=12cm]{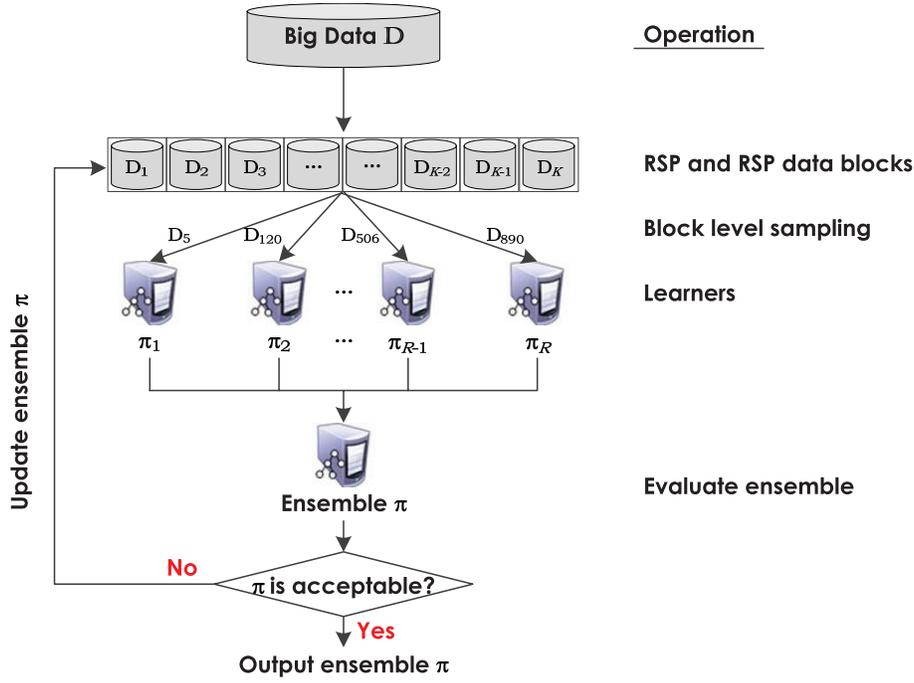}
		\caption{Asymptotic Ensemble Learning Framework: Base models are built from a sample of RSP data blocks and then combined in an ensemble model.}
		\label{fig:framework}
	\end{center}
\end{figure*}

As shown in Figure (\ref{fig:framework}), the asymptotic ensemble learning framework provides a general framework to analyze big data by building base models from randomly selected RSP data blocks of the big data set. Given a learning algorithm $\boldsymbol{F}$ and an RSP data model ${\rm{T}} = \left\{ {{{\rm{D}}_1},} \right.{{\rm{D}}_2}, \cdots ,\left. {{{\rm{D}}_K}} \right\}$, an ensemble model are learnt from RSD data blocks in batches as follows:
\begin{itemize}
	
	\item Blocks Selection: a sample of RSP data blocks is drawn without replacement from the RSP data model, e.g., ${{\rm{D}}_5}$, ${{\rm{D}}_{120}}$, ${{\rm{D}}_{506}}$ and ${{\rm{D}}_{890}}$. This sample of RSP data blocks is put in one batch. 
	
	\item Learning Base Models: A base model is built from each selected RSP data block. These base models are built in parallel on a computing cluster. For instance, four classifiers $\pi_1, \pi_2, \pi_3, \pi_4$ are built in parallel from the four selected RSP data blocks as shown in the figure.  
	
	\item Ensemble Learning and Update: The base models are collected to form an ensemble model $\Pi$ which is updated after each batch.
	\item Ensemble Evaluation: If the current ensemble model $\Pi$ is evaluated. If it does not satisfy the termination condition(s), go back to the first step; otherwise, output the ensemble model. This process continues until a satisfactory ensemble model $\Pi$ is obtained or all blocks are used up.

\end{itemize}
	
	\begin{figure}[!htb]
		\begin{center}
			\includegraphics[width=8cm,height=6cm]{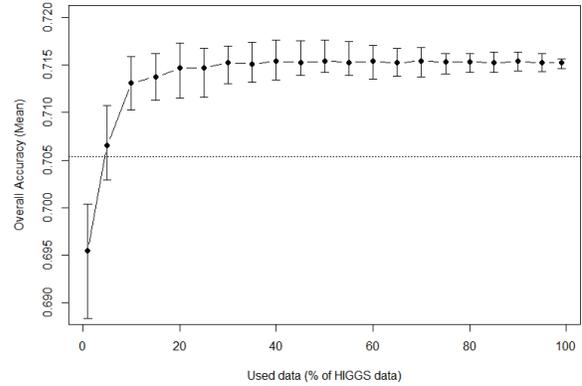}
			\caption{Accuracy of block-based ensemble classifiers on HIGGS data averaged from 100 runs of Algorithm 2. Base models were trained in batches of 5 blocks with 1\% data in each block.  The dotted line is the accuracy of a single model built using all data with the same algorithm.}
			\label{fig:higgs_accuracy}
		\end{center}
	\end{figure}	
	

The basic operations in building an ensemble model using the asymptotic ensemble learning framework are described in Algorithm \ref{alg:asymptotic_ensemble}. Base models are built from RSP data block samples until there is no significant increase in the accuracy of the ensemble model. \textit{BlocksSampling} function is used to select $g$ RSP data blocks from ${\rm{T}}$.  $\Omega()$ is an evaluation function to evaluate whether the ensemble model $\Pi$ satisfies the analysis requirements. An empirical study of this method was presented in \cite{Salloum:2016:EAA:3006299.3006306}.

A prototype of this framework was implemented using Microsoft R Server packages with Apache Spark. Taking decision trees as an example, Figure (\ref{fig:higgs_accuracy}) shows the results on HIGGS data set. Ensemble models were built using the previous ensemble process and evaluated on the same test data after each batch. We can see how the ensemble model accuracy changes after each batch. However, this change is not significant after using about 15-20\% of the data. Furthermore, the ensemble model accuracy is generally better than the accuracy of a single model built using the whole data (the dotted line in the figure). This framework can also be used to get indicators of a model's quality using only a subset of RSP data blocks. 

\begin{figure}[!htb]
	\begin{center}
		\includegraphics[width=8cm]{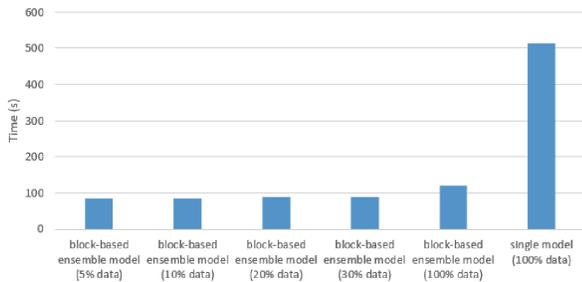}
		\caption{Training time for block-based models on a cluster of 5 nodes. Each block holds 1\% of the data (e.g., 5\% of data means that 5  blocks were used in one batch). For comparison, the time for building a single model from all data on the same cluster is shown on the far right.}
		\label{fig:higgs_time_barplot}
	\end{center}
\end{figure}

These results show that a small subset of RSP data blocks from a big data set are enough to obtain ensemble models equivalent to the models built using the whole data set. In this way, computation time can be decreased significantly as shown in Figure (\ref{fig:higgs_time_barplot}). As far as the number of RSP data blocks in a batch is equal to or smaller than the number of available executors in the computing cluster, the computation time of a batch does not vary a lot. Even when using all RSP data blocks of a big data set in one batch, which is generally not required, it still takes less time than required for training a single model from the whole data (about 8 minutes for HIGGS data set on the same cluster).

\begin{algorithm}[t]
	\caption{Asymptotic Ensemble Learning Algorithm}
	\label{alg:asymptotic_ensemble}
	\begin{algorithmic}
		\STATE {\bfseries Input:}
		
		\STATE - ${\rm{T}}$: an RSP of $\mathbb{D}$;
		
		\STATE - $f$: a learning algorithm;
		
		\STATE - $g$: number of blocks in one batch;
		\STATE {\bfseries Method:}
		\STATE ${\Pi} = \phi$;
		
		\STATE 1- {\bfseries Blocks Selection:} $S=BlocksSampling({\rm{T}},f, g)$
		
		\STATE 2- {\bfseries Base Models Learning:}\FOR {${\rm{D}}_q\in S$}			
		\STATE $\pi_q$ = $f({\rm{D}}_q)$
		\ENDFOR
		\STATE 3- {\bfseries Ensemble Update:} $\Pi$ = $\Pi$ + $\{\pi_q\}^g_{q=1}$
		\STATE 4- {\bfseries Ensemble Evaluation:}
		\STATE \textbf{If} $\Omega(\Pi)< \emph{threshold}$ (or no more blocks)
		\STATE   {Stop}
		\STATE \textbf{Else}
		\STATE   {Go to 1}
		
		\STATE {\bfseries Output: $\Pi$}

	\end{algorithmic}
\end{algorithm}

\section{Advantages and Implications}
In this work, we argue that alleviating the challenges of big data analysis, especially when data volume goes beyond the available computing resources, requires a different way of thinking. It is not efficient to treat data analysis as a pure computational problem and ignore its statistical aspects whether on the data analysis level or data management level. Applying the RSP data model on big data sets is a promising approach for a variety of data analysis tasks. It can open the door for further investigations of new innovative solutions to cope with the ever increasing volume of big data sets.

A key advantage is that RSP data blocks can be used directly as random samples without the need for expensive online sampling and extra storage space. These blocks can be used to obtain unbiased estimators of a big data set because each data block has approximately the same distribution as the whole data set. In addition, creating an RSP data model from a big data set is only performed once. After that, samples of RSP data blocks can be used directly, not only for approximate computing and ensemble learning, but also for exploratory data analysis, interactive analysis and quickly piloting models. As we can see from the experimental results, while there is no significant difference in accuracy, block-based ensembles and estimations require less computational time and resources. At least, equivalent results can be reached using only a small portion of a big data set. In this way, memory limitation is not critical anymore as data is analyzed in batches and RSP data blocks are small enough to be processed on single nodes or cores.

Since the RSP based analysis approach enables reusing sequential algorithms for conducting data analysis on RSP data blocks without the need to rewrite these algorithms for distributed and parallel environment, tackling big data analysis problems turns to solving small data problems and defining proper ensemble or combination functions. Furthermore, statistically-aware data partitioning algorithms can be implemented using the current APIs of distributed computation engines, such as Apache Hadoop and Apache Spark. As a result, many of the current algorithms can be scaled to big data without the need for new parallel implementations. This approach can also help in separating big data storage from big data analysis so that selected RSP data blocks can be loaded on or transferred to local machines or computing clusters for analysis. As such, a computing cluster can be used to analyze data sets stored in different clusters by combining the outputs from different locations. If RSP data blocks from different data centers have different probability distributions, a combination criterion can be defined to produce representative RSP data blocks of the whole data set and then analyze the combined blocks to produce the estimated results of the big data in different data centers.

\section{Conclusions}
\label{conclusions}

In this article, we introduced the random sample partition (RSP) data model to represent a big data set as a set of RSP data blocks where each RSP data block itself is a random sample of the whole data set. An RSP data model of a big data set can be created using distributed data partitioning algorithms. We showed that block level samples from an RSP data model can be used efficiently and effectively for different data analysis tasks such as statistics estimation and ensemble learning. We demonstrated how the asymptotic ensemble learning framework is used as a general framework for building block-based ensemble models using block samples from an RSP data model. For further testing and investigation, some of our future works include implementing different statistically-aware partitioning algorithms, extending our framework for building random subspace base models in order to increase the variety of base models as well as testing other data analysis tasks.

\section{Acknowledgments}
 The first author and second author contributed equally the same to this paper which was supported by grants from National Natural Science Foundation of China  (No. 61473194, 61503252) and China Postdoctoral Science Foundation (No. 2016T90799).

%

\bibliographystyle{abbrv}
\bibliography{random_partition_arxiv}  

\begin{thebibliography}{10}

\bibitem{Aggarwal:2015:DMT:2778285}
C.~C. Aggarwal.
\newblock {\em Data Mining: The Textbook}.
\newblock Springer Publishing Company, Incorporated, 2015.

\bibitem{ISI:000251994700031}
J.~Dean and S.~Ghemawat.
\newblock {Mapreduce: Simplified data processing on large clusters}.
\newblock {\em {COMMUNICATIONS OF THE ACM}}, {51}({1}):{107--113}, {JAN}
  {2008}.

\bibitem{Gantz2011}
J.~Gantz and D.~Reinsel.
\newblock Extracting value from chaos, 2011.

\bibitem{goiri2015approxhadoop}
{\'I}.~Goiri, R.~Bianchini, S.~Nagarakatte, and T.~D. Nguyen.
\newblock Approxhadoop: Bringing approximations to mapreduce frameworks.
\newblock In {\em Proceedings of the Twentieth International Conference on
  Architectural Support for Programming Languages and Operating Systems}, pages
  383--397. ACM, 2015.

\bibitem{Gretton:2012:KTT:2188385.2188410}
A.~Gretton, K.~M. Borgwardt, M.~J. Rasch, B.~Sch\"{o}lkopf, and A.~Smola.
\newblock A kernel two-sample test.
\newblock {\em J. Mach. Learn. Res.}, 13:723--773, Mar. 2012.

\bibitem{Kleiner:2012tm}
A.~Kleiner, A.~Talwalkar, P.~Sarkar, and M.~I. Jordan.
\newblock {The Big Data Bootstrap}.
\newblock In {\em ICML}, 2012.

\bibitem{Salloum2016}
S.~Salloum, R.~Dautov, X.~Chen, P.~X. Peng, and J.~Z. Huang.
\newblock Big data analytics on apache spark.
\newblock {\em International Journal of Data Science and Analytics},
  1(3):145--164, 2016.

\bibitem{Salloum:2016:EAA:3006299.3006306}
S.~Salloum, J.~Z. Huang, and Y.~He.
\newblock Empirical analysis of asymptotic ensemble learning for big data.
\newblock In {\em Proceedings of the 3rd IEEE/ACM International Conference on
  Big Data Computing, Applications and Technologies}, BDCAT '16, pages 8--17,
  New York, NY, USA, 2016. ACM.

\bibitem{hdfs2010}
K.~Shvachko, H.~Kuang, S.~Radia, and R.~Chansler.
\newblock The hadoop distributed file system.
\newblock In {\em 2010 IEEE 26th Symposium on Mass Storage Systems and
  Technologies (MSST)}, pages 1--10, May 2010.

\bibitem{export:159275}
M.~Vojnovic, F.~Xu, and J.~Zhou.
\newblock Sampling based range partition methods for big data analytics.
\newblock Technical Report MSR-TR-2012-18, February 2012.

\bibitem{Zaharia2012}
M.~Zaharia, M.~Chowdhury, T.~Das, and A.~Dave.
\newblock {Resilient distributed datasets: A fault-tolerant abstraction for
  in-memory cluster computing}.
\newblock {\em NSDI'12 Proceedings of the 9th USENIX conference on Networked
  Systems Design and Implementation}, pages 2--2, Apr. 2012.

\bibitem{Zaharia:2016:ASU:3013530.2934664}
M.~Zaharia, R.~S. Xin, P.~Wendell, T.~Das, M.~Armbrust, A.~Dave, X.~Meng,
  J.~Rosen, S.~Venkataraman, M.~J. Franklin, A.~Ghodsi, J.~Gonzalez,
  S.~Shenker, and I.~Stoica.
\newblock Apache spark: A unified engine for big data processing.
\newblock {\em Commun. ACM}, 59(11):56--65, Oct. 2016.

\end{thebibliography}

\end{document}